\documentclass[10pt]{article}
\usepackage{graphicx} 
\usepackage[left=2cm,right=2cm,top=2cm,bottom=2cm]{geometry}
\usepackage{amsmath}
\usepackage{amsfonts}
\usepackage{amssymb}
\usepackage{xcolor}
\usepackage{soul}
\usepackage{setspace}
\usepackage{authblk}
\usepackage{hyperref}
\usepackage{graphicx}
\usepackage{lmodern}
\title{Machine Learning Approach to Study of Low Energy Alpha-Deuteron Elastic Scattering using Phase Function Method}

\author[1]{Arushi Sharma}
\author[1]{Ayushi Awasthi}
\author[1]{Jyoti Sharma}
\author[1]{Ishwar Kant}
\author[2]{M. R. Ganesh Kumar}
\author[1]{O. S. K. S. Sastri}

\affil[1]{Department of Physics and Astronomical Sciences, Central University of Himachal
Pradesh, Dharamshala, 176215, Himachal Pradesh, Bharat (India)}
\affil[2]{Applied Materials India Private Limited, Bengaluru, 560066, Bharat (India)}

\date{}

\begin{document}

\maketitle
\begin{abstract}
The present study focuses on the analysis of elastic scattering in the alpha-deuteron system for low-energy data up to 14 MeV. This system holds particular significance due to its direct connection to the $^6Li$ production reaction, which is an important process in nuclear astrophysics and light nuclei synthesis. The primary objective of this investigation is to explore the low-lying excited states of positive parity with isospin T=0, which are crucial for accurately describing the resonance behavior and underlying nuclear dynamics of the system.\\

Central idea: To obtain the interaction potential using the inverse scattering method, we have employed the Physics-Informed Machine Learning (PIML) approach. In this framework, the machine learning algorithm is guided by the underlying physical laws, enabling the accurate extraction of the inverse scattering potential from the elastic scattering data.\\

Methodology: As a reference potential, a combination of three smoothly joined Morse functions has been utilized, characterized by ten model parameters. These parameters are optimized in an iterative fashion using a Genetic Algorithm to ensure the best fit to the phase shifts extracted from the experimental scattering data. 
The process of optimization is guided by the computed scattering phase shifts by solving the phase equation using 5th order Rk-method for the reference potential in each iteration \\

Results: Our approach yields inverse potentials for both single and multi
channel scattering. Using the Scattering Phase Shifts obtained from these inverse potentials, we calculate the partial cross-section to determine the resonance energies and deacy width. The obtain values of resonance energies and decay width for $^3D_1$, $^3D_2$ and $^3D_3$ states of $\alpha-d$ are in correspondence with the eperimental results.\\

Conclusion: It can be concluded that our machine learning-based approach for constructing the inverse potential offers a novel and complementary technique to existing direct methods.
\end{abstract}

\setstretch{1.5}
\section{Introduction}
The $\alpha-d$ reaction has been one of the most studied processes in recent decades, providing crucial information on the excited states of $^6Li$ \cite{hahn1985p, PhysRev.98.586, senhouse1964excited, allen1960energy, inglis1953energy}. Recently, $\alpha-d$ elastic scattering has been investigated using two potential models to determine the abundance of $^6Li$ in ancient halo stars, contributing to our understanding of primordial nucleosynthesis and stellar evolution \cite{kuterbekov2017peculiarities}. The radiative capture of $\alpha+d \to ^6Li + \gamma$ is the only process that produces $^6Li$ in the big bang model \cite{hammache2010high}. Direct measurements of this radiative capture process at astrophysically relevant energies of $(E\leq300~ keV)$ are virtually impossible because of the extremely low cross section. 
Consequently, indirect methods remain the only viable approach to obtain insight into the formation of $^6Li$. The interaction potentials required for such calculations are derived by fitting the $\alpha-d$ elastic scattering phase shifts for the S, P, and D wave contributions \cite{mukhamedzhanov2011reexamination}.\\


The direct problem involves predicting the outcome of a scattering experiment – the distribution of outgoing particles by modeling the description of the nuclear interaction potential using mathematical functions and solving the TISE to obtain wavefunction from the S-matrix that describe various scattering variables. 

In inverse scattering theory, instead of deriving the scattering S-matrix from the interaction potential between the scattering particles, we reverse the process and calculate the interaction potential directly from the S-matrix \cite{mackintosh2012inverse}. In essence, inverse scattering involves determining the interaction potential solely from the available scattering observables. Inverse scattering theory constructs a separable model potential that serves as an adequate input to calculate many-nucleon properties \cite{tabakin1969inverse}.\\
Typically, one requires all the bound state energies ($E<0$) and the scattering phase shifts at all energies E$>0$ up to $\inf$, to be able to exactly solve inverse scattering problems.
Thus, inverse scattering problems are generally more challenging and less developed than the corresponding direct methods, as they require reconstructing an unknown system from limited observed data. The process involves solving an ill-posed problem, often requiring regularization techniques and optimization methods\cite{mackintosh2012inverse}.

Recently, Muzafarov\cite{muzafarov2007two} proposed a comprehensive solution to the case of nonlocal separable potentials, providing a rather intricate algorithm that encompasses the entire class of phase equivalent potentials within a single parial wave channel. 
Several studies have shown that machine learning (ML) approaches are effective in solving inverse 
scattering problems\cite{abbasi2024physics}.

Deep neural networks that have been highly successful in uncovering the underlying relationships in the model usually need large datasets, which are typically scarce in scattering problems\cite{jing2023deep}. Instead, these networks can be trained using additional information by enforcing physical laws. Physics-informed machine learning integrates data with mathematical models, guiding solutions toward physical plausibility while improving accuracy and efficiency \cite{karniadakis2021physics}.
Researchers have proposed leveraging ML to optimize the parameters in model 
simulations and to directly tackle many-body problems by enhancing traditional Monte-Carlo
simulation methods \cite{balassa2024nonautonomous, raghavan2021machine}. \\

In this paper, we will use a genetic algorithm based global optimization to obtain the reference potential model parameters using the available experimental energy data up to $14~ MeV$. The optimization procedure utilizes the scattering phase shifts obtained by numerical solution of the phase equation for a chosen $\ell$-channel, and minimizes a loss function, which quantifies the discrepancy between the simulated and expected data. The crucial part of this procedure is the choice of reference potential whose parameters are optimized to obtain the inverse potential.\\


In charged particle scattering, the Coulomb interaction plays a significant role alongside the nuclear interaction. The phase function method involves matching spherical Bessel functions at the boundary where the interaction becomes negligible. To model the electromagnetic interaction, various screened Coulomb potentials, such as the error function (\textit{erf}) and the atomic Hulthén potential, have been employed in the literature \cite{awasthi2024comparative, sastri2022innovative}.

However, the long-range nature of the \textit{erf} function requires it to be abruptly truncated at large distances, and the atomic Hulthén potential demands fine-tuning of the screening parameter $a$, which alters the potential depth for different values of $\ell$. To address these limitations, we adopt a 
reference potential made up of three Morse function which are smoothly joined at the in between boundaries to accurately describe the short-range, medium-range, and long-range behavior of the interaction.

The range of values limited by the bounds set for the various model parameters of the reference function results in a diverse set of curves in the sample space. The optimal inverse potential, which best reproduces the 
expected phase shifts, is identified by adjusting the parameters of the reference function to minimize a cost function, such as the Mean Absolute Percentage Error (MAPE). This procedure of constructing inverse potentials by solving the phase equation using a reference function is known as the reference potential approach.\\

Based on this methodology, we have previously modeled the interaction using two Morse functions for $\alpha-\alpha$\cite{sastri2024constructing} and three Morse functions for n-p \cite{awasthi2024high} system to obtain the inverse potential across various $\ell$-channels. Here we extend the methodology to construct the inverse potential for $\alpha-d$ system to study the low-lying excited states for $^6Li$ corresponding to T =0.
 
\section{Methodology}
In this section, we discuss the significance of the reference function in inverse scattering and demonstrate the non-linear relationship between the observable and the obtained inverse potential. The phase function method (PFM) is used as a guiding principle in machine learning algorithms to determine the inverse scattering potential via the Reference Potential Approach (RPA). Hence, referred to as physics-informed machine learning (PIML). PFM provides a versatile and efficient framework for analyzing quantum scattering problems, offering distinct advantages and simplicity. We elaborate the PFM for both single and multi-channel scattering \cite{babikov1967phase} and discuss the machine-learning-based meta-heuristic algorithm used to optimize the model parameters \cite{karniadakis2021physics}.

\subsection{Reference Potential Approach}
Understanding nuclear structure, reactions, and scattering requires an understanding of the nuclear interaction potential, which describes the forces between nucleons. The nuclear interaction potential exhibits a complex structure with three distinct regions, corresponding to different physical behaviors at short, intermediate, and long ranges.\\

An effective potential encompasses all three regions: short-range repulsion, intermediate-range attraction, and long-range decay to provide an accurate description of nuclear scattering dynamics.\\
Therefore, we use a combination of Morse functions to construct the reference potential. As a sum of exponential terms, the Morse function effectively captures the complex structure of the interaction potential, including short-range repulsion and the long-decaying tail at large distances. The availability of an analytical solution for the \( l = 0 \) state makes it particularly suitable for modeling nuclear interactions that vary with distance.\\
The reference potential is defined as 
\begin{eqnarray}
U(r)= 
\begin{cases}
  U_0 = V_0 + D_0[e^{-2\alpha_0(r-r_0)}-2e^{-\alpha_0(r-r_0)}],& \text{if } r\leq X_1\\
   
    U_1 = V_1 + D_1[e^{-2\alpha_1(r-r_1)}-2e^{-\alpha_1(r-r_1)}],& \text{if } X_1<r< X_2\\
   U_2 =   V_2 - D_2[e^{-2\alpha_2(r-r_2)}-2e^{-\alpha_2(r-r_2)}],&\text{if}~ r\geq X_2
\end{cases}
\label{RPA}
\end{eqnarray} 
where $U_0$ represents the short-range interaction, $U_1$ represents the intermediate range, including the Coulomb interaction, and the third term $U_2$ used in an inverted form to account for the long decay tail of Coulomb at large distances, describes the scattering behaviour of charged systems.
$X_1$ and $X_2$ represent the boundary points for the 3 Morse function defined at three different regions.\\
To ensure the smoothness of the potential at these boundary points, the functions and their derivative must be continous at $X_1$ and $X_2$. That is:
\begin{eqnarray}
U_i|_{X_{i}}=U_{i+1}|_{X_{i}}\\
\frac{dU_i}{dr}|_{X_{i}}=\frac{dU_{i+1}}{dr}|_{X_{i}}
\label{bc}
\end{eqnarray}
where $i$ takes values 1 $\&$ 2.\\
Each Morse function consists of four parameters- V, D, $\alpha$, and 
r. Thus, the reference potential includes a total of 14 parameters, including $X_1$ and $X_2$, which need to be optimized. By applying four boundary conditions, four of these parameters can be determined, reducing the number of free model parameters to 10. Consequently, the reference model represents a family of curves defined by ten adjustable parameters. These relationships between variables maintain consistency and reduce the number of free parameters in the model. These parameters are optimized using the Physics-Informed Machine Learning (PIML) paradigm through inverse scattering, employing the variable phase approximation.\\

\subsection{Phase Function Method:}
In order to solve the inverse scattering problem, the Phase Function Method (PFM) offers a straightforward and effective method for reconstructing the unknown potential from scattering data. 
In PFM, the Schrodinger equation(a linear homogeneous equation of the second order) \cite{hans2008nuclear}, is reformulated into a first-order non-linear Riccati equation, which evolves with the radial coordinate and directly relates the scattering phase shifts to the potential \cite{zhaba2016phase}. At each point, the phase equation represents the scattering phase shift corresponding to the potential, providing insight into the influence of different regions of the potential.\\
The Schrodinger wave equation for a spinless particle with energy E and orbital angular momentum $\ell$ undergoing scattering is given by 
\begin{equation}
\frac{\hbar^2}{2\mu}\left[\frac{d^2}{dr^2} + \left(k^2-\frac{\ell(\ell+1)}{r^2}\right)\right]u_{\ell}(k,r) = V(r) u_{\ell}(k,r)
\label{SchrEq}
\end{equation}
where  $\mu$ is the reduced mass of the system.
\\ 
The projectile energy in lab frame, $E_{lab}$, would be related to centre-of-mass energy $E_{cm}$ using standard relation

\begin{equation}
    E_{cm} = \frac{m_{T}}{m_{T}+m_{P}}E_{\ell ab} .
\end{equation}
Here, $m_{T}$ and $m_{P}$ are masses of target and projectile respectively.\\

The central idea of the Variable Phase Approach (VPA) is to obtain the phase shift 
 $\delta$  directly from physical quantities, such as the interaction potential, without solving the Time-Independent Schrödinger Equation (TISE) for wave functions 
 $u(r)$, which are typically used to determine 
$\delta_\ell(k,r)$. The PFM was originally developed for the case of scattering by a spherically symmetric potential, but it was later expanded to include more generic conditions such as scattering in the field of many channels, relativistic equations, non-central forces, single channel, multi-channel and so forth.

\subsubsection{Single Channel scattering:}
For elastic scattering of different channels by a central potential where different orbital angular momenta remain independent, the second-order Time Independent Schrodinger Equation (TISE) is transformed into a Riccati-type equation given by\cite{sastri2022innovative}
\begin{equation}
\frac{d\delta_\ell(k,r)}{dr}=-\frac{U(r)}{k}\bigg[\cos(\delta_\ell(k,r))\hat{j}_{\ell}(kr)-\sin(\delta_\ell(k,r))\hat{\eta}_{\ell}(kr)\bigg]^2
\label{PFMeqn}
\end{equation}
where $U(r)=\frac{2\mu V(r)}{\hbar^2}$ and $\delta_\ell(k,r)$ is called as phase function. The initial condition for the phase equation is $\delta_\ell(0)=0$, which corresponds to the actual absence of potential at $r=0$. Here $\hat{j_\ell}(k,r)$ and $\hat{n_\ell}(k,r)$ are the Riccati-Bessel and Riccati-Neumann functions\cite{khachi2023inverse}.

The phase shift $\delta_\ell(k,r)$ can be seen as a real function of k and characterizes the strength of scattering of any partial wave that is say $\ell^{th}$ partial wave of the potential U(r). The riccati-Hankel function of the first kind relates to the Riccati-Bessel Function $\hat{j_\ell}(k,r)$ and the Riccati-Neumann function $\hat{n_\ell}(k,r)$ as $\hat{h}_{\ell}(r)=-\hat{\eta}_{\ell}(r)+\textit{i}~ \hat{j}_{\ell}(r)$. The Riccati-Bessel and Riccati-Neumann functions can be derived using the recurrence relations.
\begin{equation}
   {\hat{j}_{\ell+1}}(kr)=\frac{2\ell+1}{kr} \hat{j_\ell}(kr)-{\hat{j}_{\ell-1}}(kr)
    \label{R1}
\end{equation}
\begin{equation}
   {\hat{\eta}_{\ell+1}}(kr)=\frac{2\ell+1}{kr} \hat{\eta_\ell}(kr)-{\hat{\eta}_{\ell-1}}(kr)
   \label{R2}
\end{equation}

So phase equation, for $\ell$=0 is:
\begin{equation}
    \delta_0'(k,r)=-\frac{U(r)}{k}\sin^2[kr+\delta_0(r)]
    \label{NS}
\end{equation} 

for $\ell$=1 is
\begin{equation}
\delta_1'(k,r)=-\frac{U(r)}{k}\bigg[\frac{\sin(\delta_1+(kr))-(kr) \cos(\delta_1+(kr))}{(kr)}\bigg]^2
\end{equation}

and for $\ell$=2 is
\begin{equation}
\delta_2'(k,r) = -\frac{U(r)}{k}\bigg[-\sin{\left(\delta_2+(kr) \right)}-\frac{3 \cos{\left(\delta_2 +(kr)\right)}}{(kr)} + \frac{3 \sin{\left(\delta_2 + (kr) \right)}}{(kr)^2}\bigg]^2 
\label{NS3}
\end{equation}

$\delta_\ell(k,r)$ is known as the phase function and its value at r=R provides the phase shift corresponding to the interaction potential U(r) at that point. In the asymptotic limit $r \to\infty$, the phase shift values becomes constant. Eqn.\ref{PFMeqn} is non linear equation and can be solved numerically using Rk-5 method with initial condition $\delta_\ell(k,0)=0$

\subsubsection{Multi-channel Scattering}
For the non-central tensor interaction and many-channel inelastic scattering, PFM can be expanded. Tensor interaction is mainly considered where the elastic scattering of integral spin nucleons is taken into account. In the triplet spin state, there is mixing of partial waves for a total angular momentum J due to tensor forces correspond to different orbital angular momenta $L=J\pm1$. Equations for the coupled radial wave function are given as \cite{babikov1964some}:
\begin{equation}
\frac{d^{2}u_J(k,r)}{dr^2} +\left(k^2 - \frac{J(J-1)}{r^2} - V_{J,J-1}\right) u_{J}(k,r) - T_{j}w_{J}(k,r) = 0
\label{couple}
\end{equation}
\begin{equation}
\frac{d^{2}w_J(k,r)}{dr^2} +\left(k^2 - \frac{(J+2)(J+1)}{r^2} - V_{J,J+1}\right) w_{J}(k,r) - T_{j}u_{J}(k,r) = 0
\label{couplea}
\end{equation} 

Coupling of these equations complicates the calculation due to the involvement of two phase shifts and mixing component, making it difficult to calculate the scattering parameters. Coupling of the waves makes it challenging to extract the slowly increasing solution for small r, since one solution dominates the other. PFM allows us to drive the set  of first-order non-linear equations for three functions which is free from this disadvantage. \\
Equations of the PFM for various representation of the parameters have been derived by kynch\cite{kynch1952two}, babikov \cite{babikov1964some} and Cox $\&$ perlmutter\cite{cox1965method}.
Here, in this paper, we consider the equations for the functions $\delta_{J, J-1}(r)$, $\delta_{J, J+1}(r)$ and $\epsilon_{J}(r)$ which are associated with the ``Stapp parametrization", widely employed in nuclear physics \cite{babikov1964some,zhaba2019stapp}. The equations for Stapp parameterization can be written for a particular J as:
\begin{equation}
\begin{split}
\frac{d\delta_{J,J-1}}{dr} = & \frac{-1}{k\cos2\epsilon_J}\Bigg[V_{J,J-1} \left(\cos^{4}\epsilon_J P^{2}_{J,J-1} - \sin^{4}\epsilon_J Q^{2}_{J,J-1}\right) \\
& - V_{J,J+1} \sin^{2}\epsilon_{J}\cos^{2}\epsilon_{J}\left(P^{2}_{J,J+1}-Q^{2}_{J,J+1}\right) \\
& - 2T_{J}\sin \epsilon_{J} \cos \epsilon_{J}\left(\cos ^{2} \epsilon_J P_{J,J-1}Q_{J,J+1} - \sin ^{2}{e_J} P_{J,J+1}Q_{J,J-1}\right)\Bigg]
\end{split}
\label{J-1}
\end{equation}
\begin{equation}
\begin{split}
\frac{d\delta_{J,J+1}}{dr} = & \frac{-1}{k\cos2\epsilon_J}\Bigg[V_{J,J+1} \left(\cos^{4}\epsilon_J P^{2}_{J,J+1} - \sin^{4}\epsilon_J Q^{2}_{J,J+1}\right) \\
& - V_{J,J-1} \sin^{2}\epsilon_{J}\cos^{2}\epsilon_{J}\left(P^{2}_{J,J-1}-Q^{2}_{J,J-1}\right) \\
& - 2T_{J}\sin \epsilon_{J} \cos \epsilon_{J}\left(\cos ^{2} \epsilon_J P_{J,J+1}Q_{J,J-1} - \sin ^{2}{\epsilon_J} P_{J,J-1}Q_{J,J+1}\right)\Bigg]
\end{split}
\end{equation}
\label{J+1}
\begin{equation}
\begin{split}
\frac{d \epsilon_J}{dr} = & \frac{-1}{k}\Bigg[T_{J} \left(\cos^{2}\epsilon_J P_{J,J-1}P_{J,J+1} + \sin^{2}\epsilon_J Q_{J,J-1}Q_{J,J+1}\right) \\
& - V_{J,J-1} \sin \epsilon_{J}\cos \epsilon_{J} P_{J,J-1} Q_{J,J-1} - V_{J,J+1} \sin \epsilon_J \cos \epsilon_J P_{J,J+1} Q_{J,J+1} \Bigg]
\end{split}
\label{e}
\end{equation}
where $P_{J,\ell} (r)$ and $Q_{J,\ell} (r)$ can be defined as
\begin{align*}
P_{J,\ell}(r) &= \cos(\delta_{J,\ell}(r)) \hat{j}_{\ell}(kr) - \sin(\delta_{J,\ell}(r)) \hat{\eta}_{\ell}(kr) \\
Q_{J,\ell}(r) &= \sin(\delta_{J,\ell}(r)) \hat{j}_{\ell}(kr) + \cos(\delta_{J,\ell}(r)) \hat{\eta}_{\ell}(kr)
\end{align*}
Eq. \ref{J-1}- \ref{e} are three non-linear coupled first-order equations which we can be solved using RK-5th order with initial condition $\delta_{J, J-1}(0) = 0, ~~~ \delta_{J, J+1}(0) = 0 ~~$and$ ~~~ \epsilon_{J}(0) = 0$.
 The non-linear differential equation is numerically integrated from the origin to the asymptotic region, allowing the direct calculation of the scattering phase shift for different projectile energies in the laboratory frame.
A family of smooth curves is considered as reference input to the phase equation to obtain the best-optimized solution, which produces the correct scattering phase shifts and ensures the physical validity of the interaction model. In the $\alpha$-deuteron scattering system, mixing of channels occurs only for J=1 state at energies up to $14~MeV$. This means that the total angular momentum of the system allows for a coupling between $^3S_1$ and $^3D_1$ states. These two states form the coupled channel in the Scattering process. The parameter $\epsilon$ quantifies the degree of mixing between these channels, describing how the wavefunction transitions between the S-wave and D-wave components due to the tensor force in the interaction. This mixing plays a crucial role in understanding the scattering dynamics and the properties of the system.

 \subsection{Cross section}
  From the Scattering Phase Shifts (SPS) $\delta_\ell(E)$ of each orbital angular momentum $\ell$, one can calculate the partial Scattering Cross Section (SCS) $\sigma_\ell(E)$ using the formula:

\begin{equation}
\sigma_l(E;S,J)=\frac{4\pi}{k^2}\left((2\ell+1)~\sin^2(\delta_\ell(E;S,J))\right)
\label{a}
\end{equation}
 This equation enables the calculation of partial scattering cross-sections from the extracted scattering phase shifts. The scattering phase shifts, which describe the modification of the wave function due to the interaction potential, are directly related to the cross-section through partial wave analysis. By plotting the energy $E$ as a function of the partial cross-section, the resonance energies corresponding to the $^3D_1$, $^3D_2$, and $^3D_3$ states of the alpha-deuteron system can be identified. This method provides a systematic approach for analyzing the energy dependence of the scattering process and extracting precise information about the resonance characteristics of the system.

\section{Algorithm for Physics Guided Parameter Optimization}
Optimization involves iteratively adjusting a model's parameters to minimize or maximize a predefined objective function, such as the loss function in supervised learning. The goal of optimization is to find the parameters that best fit the given data while generalizing effectively to unseen data, thereby improving prediction accuracy and overall model performance.\\

Optimization algorithms \cite{koziel2011computational} enable models to learn from data, adapt to complex patterns, and produce accurate predictions by adjusting model parameters. The choice of an optimization method significantly affects the efficiency and convergence of a machine learning model, making it a crucial decision during the model development process. \\
To construct the inverse potential directly from the available experimental data, we optimize the parameters of the reference potential using a meta-heuristic algorithm inspired by natural selection and evolution. The machine-learning-based genetic algorithm iteratively refines parameter sets to minimize the error between predicted and experimental values by reducing the cost function\cite{man1996genetic}. It begins with a population of feasible solutions and evolves to new solution sets at each iteration, aiming to improve fitness and achieve optimal performance.\\
Steps:
\begin{itemize}
    \item Initialisation: Arrays of projectile energies $E_{lab}$ in the laboratory frame and the corresponding scattering phase shifts $\delta(E)$ are given as input for analysis. The variation of scattering phase shifts $\delta(E)$ as a function of energy $E$ provides insight into emergent resonances through the analysis of the slope's behavior. Ranges are defined for the model parameters of the reference potential to narrow the optimization space and focus on a specific region. To improve convergence, previously obtained parameter sets can be included in the initial population, using past solutions to make the search for the global optimum more efficient.
    
    \item Genetic Algorithm\cite{katoch2021review}: The Process starts with the creation of a random population of parameters that represent possible solutions of the optimization routine. Each generation involves selecting of candidate solution based on their fitness. Their fitness is determined based on the cost function, i.e. Mean Absolute Percentage Error. MAPE between the simulated and experimental SPS is computed as
    \begin{equation}
        MAPE= \frac{1}{N}\sum_{i=1}^N |\frac{\delta_i^{exp}-\delta_i^{sim}}{\delta_i^{exp}}|\times100
    \end{equation}
    The solution with lower MAPE values are selected for reproduction, where genetic operators such as crossover and mutation create new offspring. This process of selection, cross-over, and mutation continues over multiple generations, refining the model parameters to minimize the cost function and improve the fit to the experimental data.
    \item Potential Determination: The reference potentials for each parameter produced by the parent class are calculated. Following this, the parameter ranges are adjusted to ensure that the resulting inverse potential aligns with physical constraints and meets the specified conditions. Within this framework, our machine-learning-based heuristic algorithm employs the phase equation that governs the scattering process to optimize and adjust the potential parameters.
    \item Numerical Solution of Phase Equation:  To compute the simulated
 scattering phase shifts (SPS), denoted as $\delta_l$  one solves the phase equation
 numerically using the fifth-order Runge-Kutta (RK-5) method. This solution uses the initialized reference potential as input, providing a simulated scattering phase
 shifts that can be compared with the expected data obtained from experimental
 cross-sections
    \item Testing and validation: The optimization procedure continues until the mean absolute percentage error (MAPE) does not change anymore, indicating that the method has successfully identified the ideal model parameters and converges to an optimal solution.
Once the optimization is complete, the inverse potential is successfully constructed with the best optimized parameters. To ensure reproducibility, the process can be repeated using different initial seeds.
Exploring different parameter ranges is crucial to avoid missing a solution with a lower MAPE and to ensure the accuracy of the results.

\end{itemize}

\section{Results}
\subsection{Database}
The scattering phase shift (SPS) data for various $\ell-$channels has been taken from schmelzbach et al.\cite{schmelzbach1972phase} ($E_d = 3.0~ to ~5.8~MeV$) and Jenny et al.\cite{jenny1983phase} ($E_d = 6~ to~ 14~ MeV$).
A significant portion of the experimental analysis relies on analyzing power, which is subject to substantial errors arising from fluctuating systematic uncertainties, including background subtraction uncertainty, beam position stability, and spin-angle uncertainty. To accurately model the resonance observed in low-energy experimental data for deuteron-alpha elastic scattering (below $3~ MeV$ CM energy), D-wave effects must be included in the theoretical cross-section calculations. This necessitates the analysis of low-energy D-wave data.
The two-body $\alpha-d$ system can have the following states, which correspond to the orbital angular momentum $\ell=0,~1~\&~2$, based on the spin-zero of alpha and spin-one of the deutron and the $\ell-$wave components of the $\alpha-d$ system: $^3S_1,~ ^3P_0, ~^3P_1,~ ^3P_2,~ ^3D_1,~ ^3D_2,~ and~ ^3D_3$. For $J=1$, $\alpha-d$ scattering allows for the coupling of partial waves with angular momentum difference of two units.

\subsection{Optimization routine to obtain Inverse Potential}

The model parameters of the reference function must be optimized in order to determine the interaction potential for the $\ell-$channels of $\alpha-d$.
The reference potential consists of three Morse functions, with a total of 14 parameters to optimize, including the boundary points $X_1$ and $X_2$. Using the four boundary conditions given in Eqn. \ref{bc}, four of these parameters can be expressed in terms of the other parameters. As a result, we were left with only ten parameters to optimize. Thus, a 10-dimensional parameter space is formed, resulting in a vast range of potential curves.
We optimized the parameters  $\alpha_0,~ \alpha_1,~ \alpha_2, r_0,~ r_1,~ r_2, ~V_2, ~D_0$.
 Meanwhile, the parameters $V_1,~ V_2, ~D_1, ~D_2$ are determined by enforcing the boundary conditions. The optimization process is carried out using a genetic algorithm, which evolves solutions through natural selection, incorporating crossover and mutation to explore the parameter space effectively.\\

The optimization routine relies heavily on the selection of bounds. The final integration distance, or $r_f$, has been set to a high value so that potential eventually becomes zero, and the boundaries are initially selected to span a broad range.

The initial set of model parameters is created at random from the interval bounds specified for each parameter, beginning with a random seed.  This will create a family of curves, resulting in an increase in MAPE and convergence time to the best solution.

  After a thousand iterations and thorough observations of the trend of phase shifts and potential, we can decrease the sample space for the parameters so that computation time for parameter optimization is reduced while the error between expected and computed phase shifts is lowered.

  The advantage of the Genetic algorithm is that it scans across a large no of possible curves while tending to converge towards the final potential.

  The optimal model parameters for single channels are shown in the Table \ref{sc}.  During optimization, it was discovered that the values of parameter $V_2$ approach zero or are in the order of $10^{-5}$. As a result, we have omitted the value of $V_2$ from the table because it consistently tends to zero across all channels.

\begin{figure}
    \centering
    \includegraphics[scale=0.4]{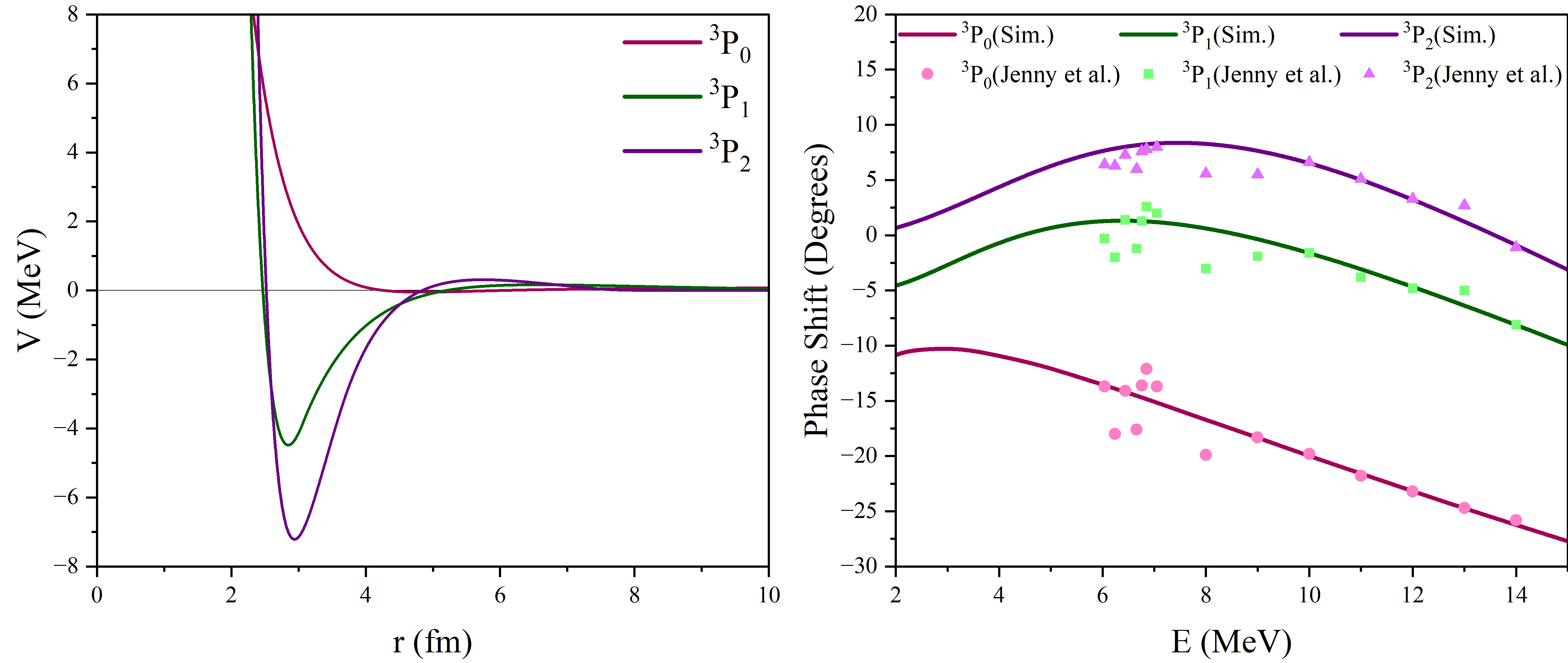}
    \caption{ Inverse potentials along with scattering phase shifts for the single channel scattering
 for P wave}
    \label{p}
\end{figure}
\begin{figure}
    \centering
    \includegraphics[scale=0.4]{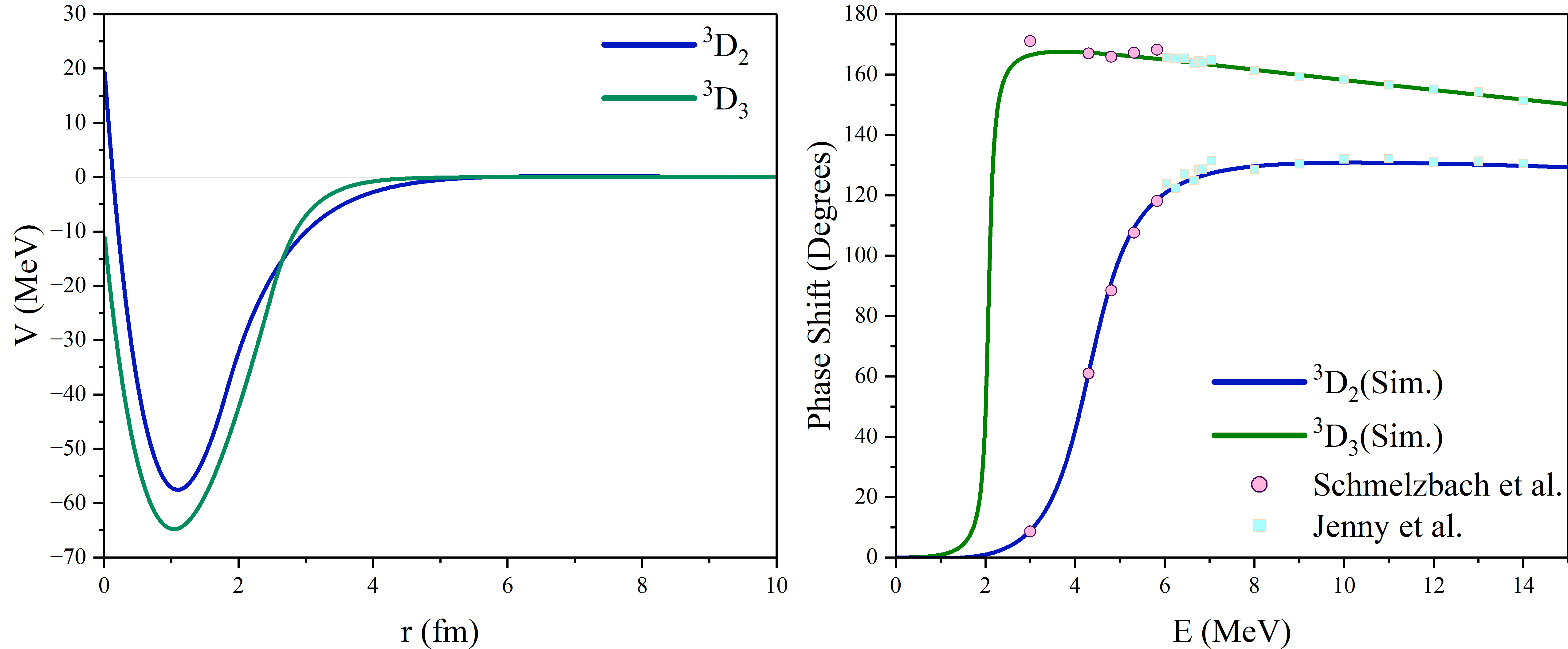}

    \caption{Inverse potentials along with scattering phase shifts for the single channel scattering
 for D wave}
    \label{d}
\end{figure}

  The obtained inverse potential using these optimized parameters and corresponding phase shifts are plotted in Fig.(\ref{p}, \ref{d})   for single $\ell-$channels.

  For the P state, data is available from 6 $MeV$ and consists of several disparate values, as shown in the Fig.\ref{p}.   The generated interaction potential represents the best feasible optimal solution. The quality of the fits is substantially energy dependent, therefore it tries to match the largest phase shift possible. 
   
  In the case of $^3P_0$, the data shows a discontinuous pattern at 6- 8 $MeV$. Beyond 8 $MeV$, the obtained SPS corresponds well to the expected data. Because phase shifts exhibit a declining trend with negative values, the potential curve can be seen to be repulsive.

   For the $^3P_1$ state, the phase shift initially increases before gradually decreasing to negative values. This slight initial rise in phase shift results in a shallow potential well with a depth of approximately 4.5 $MeV$.

Likewise, for the $^3P_2$ state, a potential well with a depth of 7.21 $MeV$ is observed. Additionally, the presence of a Coulomb barrier arises due to the repulsive interaction, which corresponds to the characteristic increasing and decreasing behavior of the phase shift.

  For the D-state, data has been obtained from Schmelzbach et al. for the energy range of 3 to 5.8 $MeV$ and from Jenny et al. for 6 to 14 $MeV$. To accurately capture the resonances in $^3D_2$ and 
$^3D_3$, it is essential to consider the low-energy data from 3 to 6 $MeV$, as the resonances occur within this range.
With the exception of the 6–8 $MeV$ region, where there is a small disagreement due to discontinuity in the expected data, $^3D_2$ phase shifts match the experimental data quite well. The phase shifts remain consistently positive, indicating an attractive interaction. The obtained inverse potential for $^3D_2$ is purely attractive, with a depth of  $V_d=57.54~ MeV$ at a distance of $r_d = 1.1~ fm$.

 For $^3D_3$, phase shifts decrease from 3 to 14 $MeV$, while data below 3 $MeV$ is extrapolated using the Genetic Algorithm. As a result of this precise extrapolation of the measured phase shifts, the resonance obtained in $^3D_3$ matches the experimental ones. This demonstrates how efficient, adaptable, and stable the algorithm is. As a result, the generated inverse potential for $^3D_3$ exhibits a depth of potential $V_d$ of 69.83 $MeV$ at a distance of $0.94~ fm$, as illustrated in Fig. \ref{d}.

   \textbf{Multi channel scattering:} 
   For $\alpha-d$ scattering up to $14 ~MeV$, coupling exist only for $J = 1$. The interaction between the two states within this channel is governed by the degree of coupling, represented by the mixing parameter $\epsilon$ \cite{jenny1983phase}. To accurately determine the phase shifts and interaction potential in multi-channel scattering, three coupled nonlinear differential equations must be solved, incorporating the mixing parameter through Stapp parametrization.

   To obtain the inverse potential for individual states and their corresponding mixing parameter, we optimized 30 parameters of the reference function. This results in a complex 30-dimensional parameter space, leading to increased computational time and challenges in adjusting the bounds for each parameter.

Initially, the phase equation is solved for single-channel scattering to estimate the bounds within which the possible solution must lie. Once a rough estimation of these bounds is obtained, it becomes easier to iteratively solve the coupled phase equation for multi-channel scattering. This approach allows for readjusting the bounds to refine the solution more effectively.

\begin{figure}
    \centering
    \includegraphics[scale=0.26]{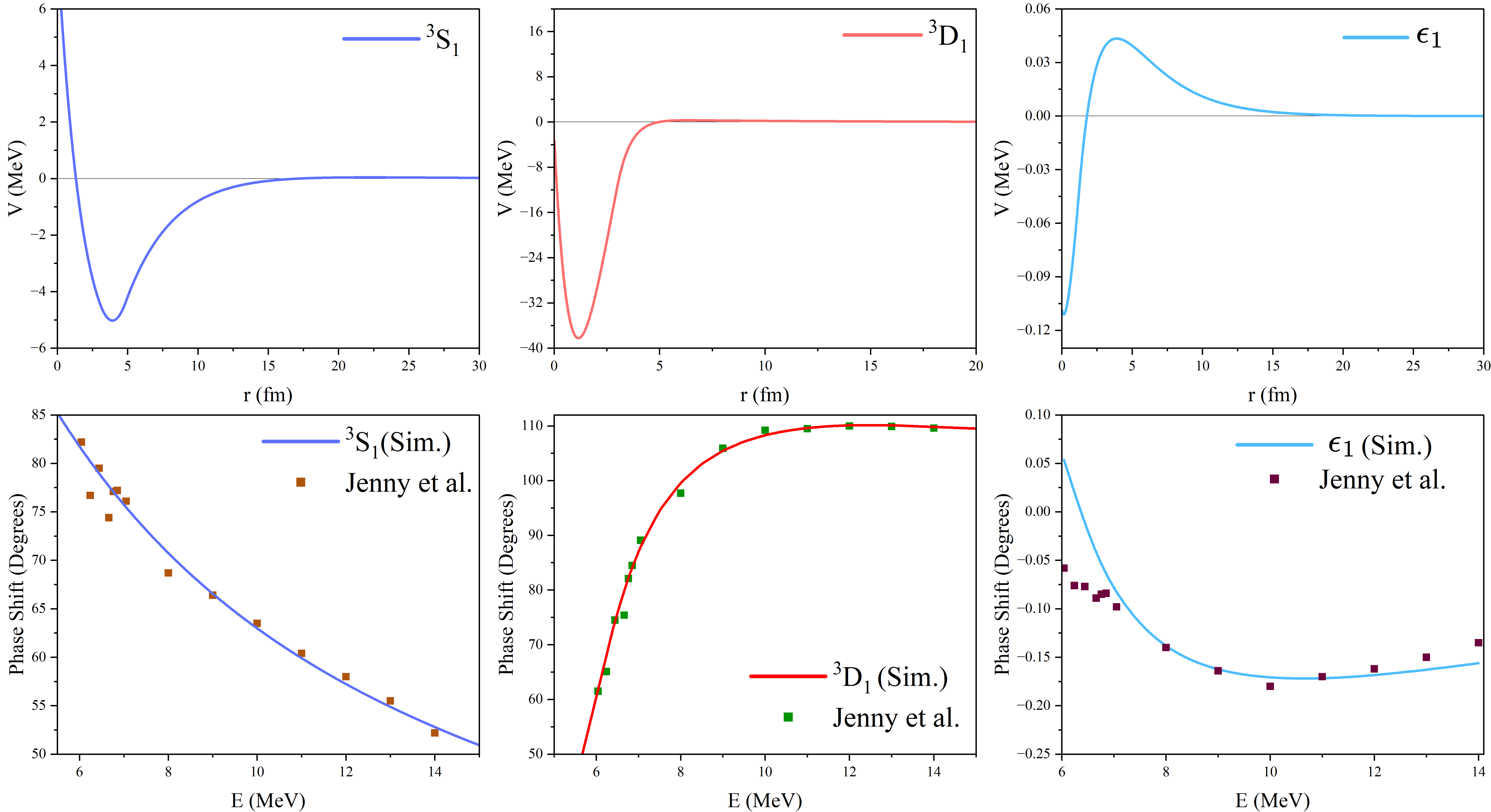}

    \caption{Inverse potentials along with scattering phase shifts for the multi-channel scattering
 of J = 1 }
    \label{mix}
\end{figure}

   The optimised model parameters for multi-channel scattering are given in Table \ref{mc}. The constructed inverse potential, along with their corresponding phase shifts, are depicted in Fig. \ref{mix}.

  For the S-D channel, the obtained phase shifts closely agree with the expected values, while discrepancies appear in the mixing channel. This is because the phase shifts for the S and D channels are significantly larger than those arising from their coupling. The deviation in the mixing channel phase shifts is primarily due to the discrete nature of the experimental data in the $6–7~ MeV$ range. However, in the $8–12~ MeV$ range, the obtained phase shifts match the expected values exactly for both the single-channel and mixing-channel cases.

   The phase shift values for the $^3S_1$ state follows a decreasing trend, indicating both repulsive and attractive characteristics, with a potential depth of 5 $MeV$ at a distance of $r_d = 4~ fm$..

For the $^3D_1$ state, the phase shift values increase positively, resulting in an inverse potential that also exhibits both repulsive and attractive features, with a depth of $V_d= 38.22~ Mev$ at a distance of $r_d = 1.15~ fm$

The mixing parameter $\epsilon$ takes negative values in an increasing order, suggesting an attractive interaction up to 4 $MeV$, beyond which repulsion becomes dominant. Consequently, the tensor potential exhibits a stronger repulsive nature in correspondence with $\epsilon$.
This type of coupling leads to positive parity states and explains the absence of strong negative parity states in $^6Li$. Supporting the interpretation of $^6Li$ as a weakly bound alpha-deuteron system with well-defined cluster dynamics.

These findings highlight the intricate balance between attractive and repulsive forces in shaping the interaction potential.

\subsection{Cross-section and Resonance energy}

\begin{figure}
    \centering
    \includegraphics[scale=0.26]{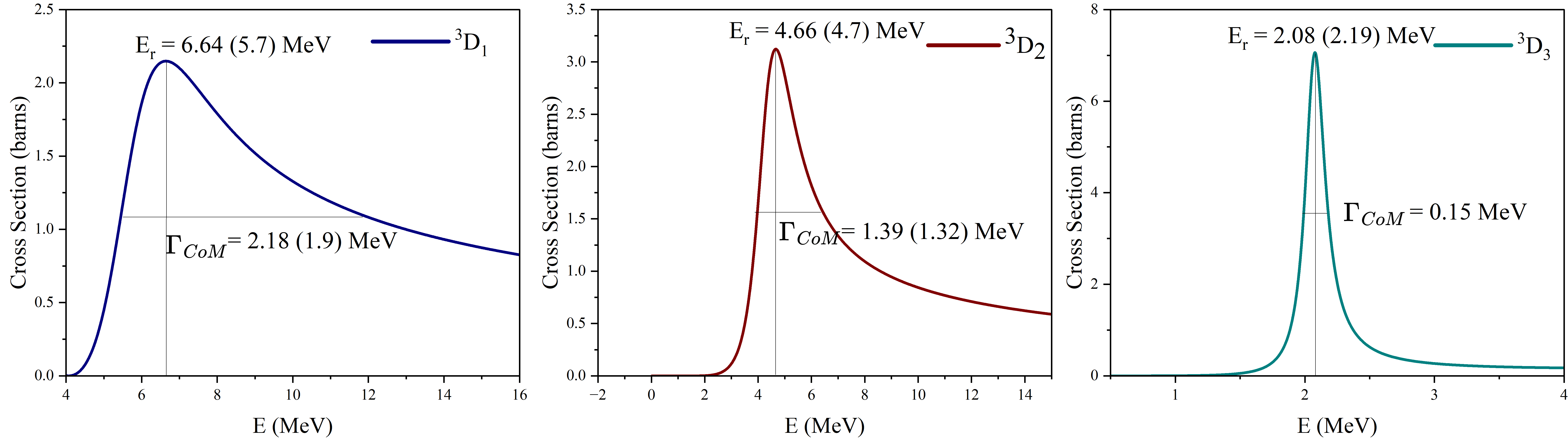}

    \caption{Partial cross section for $^3D_1,~ ^3D_2~\&~ ^3D_3$ state along with their resonance energies.}
    \label{cs}
\end{figure}

Partial cross-sections are calculated using the phase shifts corresponding to orbital angular momentum $\ell-$ channels according to the  formula
\begin{equation}
\sigma_l=\frac{4\pi}{k^2}(2l+1)sin^2\delta_l(E)
\end{equation}
The partial cross sections for resonance states $^3D_1$, $^3D_2$ and $^3D_3$ are calculated from the obtained phase shift values and plotted in Fig. \ref{cs} as a function of E ($MeV$)
 $^3D_1$ and $^3D_2$ have broad resonance while $^3D_3$ exhibits a very sharp resonance. Resonance parameters for $^3D_1$, $^3D_2$ and $^3D_3$ are given in Table \ref{csp}.

For $T=0,~~ J^{\pi}=1^+, 2^+,3^+$ are excited state of $\alpha-d$ system, which exhibits resonance below the energy range$\leq15Mev$.

The resonance energy and decay width obtained for the $^3D_1$ state is $E_r=6.64~ MeV$ and $\Gamma_{CoM}=2.18~ MeV$, while the experimental value is 5.7 $MeV$ and 1.9 $MeV$ \cite{tilley2002energy}, respectively. This slight discrepancy arises due to errors in the phase shift data. However, according to Ref.\cite{jenny1983phase}, the resonance energy for the 
$^3D_1$ state is expected to fall within the range of 4–7 $MeV$, indicating that the observed resonance is consistent with theoretical expectations.

For $^3D_2$ resonance energy obtained from the calculation of cross section from the phase shift is $E_r = 4.66 MeV$, while the experimental value is 4.36 $MeV$. The calculated decay width in centre of mass frame is $\Gamma_{CoM} = 1.39~ MeV$ while the experimental value is $\Gamma_{CoM} = 1.32~ MeV$ \cite{tilley2002energy}. This shows a close agreement between the obtained and experimental values.

Similarly, for the $^3D_3$ state, the observed resonance energy is 2.08 $MeV$, compared to the experimental value of 2.18 $MeV$. This close agreement between the calculated and experimental resonance energies highlights the accuracy of the model. The correct determination of the resonance energy for 
$^3D_3$ is a direct result of accurately predicted phase shifts at low energies. Therefore, selecting appropriate potential parameters within specific bounds is crucial. This demonstrates the algorithm's capability, reliability, and predictive power in modeling nuclear interaction dynamics and explaining resonance properties.

\section{Discussion}
In this work, we have investigated low-energy alpha-deuteron elastic scattering using a physics-informed machine learning paradigm. We obtain the scattering potential using an inverse method by incorporating physical laws into the machine learning framework. An evolutionary genetic algorithm is employed to optimize the model parameters of the reference potential. The resulting interaction potential represents the best-fit solution to the available phase shift data up to 14 $MeV$.

We construct the reference potential as a combination of three smoothly joined Morse functions to represent all possible nuclear interaction and the long range coulomb interaction. The traditional Ansatz often rely on different approximations to model the long-range Coulomb barrier. By employing the reference potential approach together with the phase function method, we are able to obtain a highly optimized inverse scattering potential that accurately reproduces the scattering phase shifts, demonstrating excellent predictive capability.

The resonance energies calculated from the cross sections, derived using the obtained phase shifts, show excellent agreement with the experimental values of the low-lying excited states of $^6Li$.

Due to the unavailability of expected phase shift data for the $^3D_3$ partial wave below 3 $MeV$, precise measurement of the appropriate resonance parameters in this region is currently impossible. The implementation of Physics-Informed Machine Learning (PIML), however, facilitates reliable extrapolation and prediction of phase shifts in this low-energy domain. Consequently, the resonance energies and decay widths for the $^3D_1$ , $^3D_2$ , and $^3D_3$  states have been determined and found to be in good agreement with the available experimental data.

Coupling in the $^3S_1$ and $^3D_1$ states indicates that the nuclear structure of $^6Li$ is mainly shaped by a cluster arrangement consisting of an alpha and a deuteron.
 The major importance of tensor interaction in binding the system is highlighted by the dominating D-Wave (L=2), pointing to a non-spherical component in nuclear shape.

 Therefore, this methodology of constructing an inverse scattering potential based on piece wise smooth Morse function as a reference potential and employing PFM with a genetic algorithm for optimization is successful in explaining the experimental outcomes of alpha-deuteron scattering.

 \section{Conclusion}
 \begin{itemize}
     \item This study confirms the dominant characteristics of $\ell=2$ wave in alpha-deutron scattering system of $^6Li$, supporting the idea that its structure is best described as an alpha core($^4He$)$+$2 valence nucleons($1p+1n$) in a p-state. The coupling of these valence nucleons to the core leads to $\ell=2$ (D-wave) and S=1(triplet state), resulting in triplet D-state $^3D_1$, $^3D_2$, $^3D_3$.
     \item The proposed methodology effectively overcomes the longstanding challenge of simultaneously incorporating long-range Coulomb interactions and short-range nuclear forces in the construction of a physically accurate scattering potential. By embedding the Coulomb effects directly into the reference potential, it avoids the need for separate treatments or approximations, resulting in a more accurate and physically consistent representation of the interaction dynamics.
     \item The main advantage of this approach is the direct determination of the Scattering Phase Shifts (SPS) using the Phase Function Method (PFM), where the interaction potential appears as a multiplicative function. This eliminates the need to solve for the full wave function, significantly simplifying the computational process while maintaining accuracy.
     \item The inverse scattering theory, when implemented computationally via the reference potential approach along with the phase function method, is effectively equivalent to the Physics-Informed Machine Learning (PIML) paradigm. In this framework, the underlying physical laws guide the machine learning algorithm to optimize the parameters of the reference potential. This enables the simulation of a broad spectrum of potential shapes across the solution space, from which the algorithm systematically converges to the optimal model potential that best reproduces the scattering data.
     \item Predictive power of genetic algorithms comes from their inherent flexibility in handling diverse data types and model structures, as well as their capacity to perform global searches across complex solution spaces rather than getting stuck in local optima. Because of its ability to adaptively refine solutions, we can predict the data at low energy and calculate the accurate resonance energy and decay width through the cross-section calculations.
     \item The observed decay widths are consistent with the wigner limit, indicating that the triplet D states have a strong spatial overlap with the alpha-deuteron cluster configuration. Additionally, the absence of strong negative parity suggests that the dominant structure of $^6Li$ is a positive parity D-wave state.
     
 \end{itemize}
Overall, the findings provide significant insight into the nuclear structure and clustering effects in $^6Li$, contributing to a deeper understanding of light nuclei and their interaction dynamics.
Our findings have substantial implications since the resulting Coulomb barrier can be used directly in low-energy S-factor calculations without the requirement to approximate the contribution of the Coulomb interaction. These aspects will be addressed separately. The alpha-deuteron scattering and radiative capture can also be studied by the Reference Potential Approach for the higher energy region.

\begin{table}[!ht]
    \centering
    \caption{ Optimized model parameters for channels exhibiting single-channel scattering}
    \resizebox{\textwidth}{!}{
    \begin{tabular}{cccccccccc}
    \hline
        \textbf{States/Parameters} & \textbf{$\alpha_0$} & \textbf{$\alpha_1$} & \textbf{$\alpha_2$} & \textbf{$r_0$} & \textbf{$r_1$} & \textbf{$r_2$} & \textbf{$X_1$} & \textbf{$X_2$} & \textbf{$D_0$} \\ \hline
        \textbf{$^3P_0$} & 2.090 & 0.976 & 0.060 & 0.750 & 4.907 & 20.370 & 0.423 & 5.279 & 93.076 \\ 
        \textbf{$^3P_1$} & 0.960 & 1.377 & 0.508 & 2.152 & 2.848 & 6.567 & 0.687 & 3.058 & 396.578 \\ 
        \textbf{$^3P_2$} & 1.531 & 0.574 & 1.887 & 2.942 & 5.744 & 11.613 & 3.636 & 7.460 & 8.834 \\ 
        \textbf{} & ~ & ~ & ~ & ~ & ~ & ~ & ~ & ~ & ~ \\ 
        \textbf{$^3D_2$} & 0.346 & 0.518 & 0.527 & 1.097 & 6.947 & 14.732 & 1.823 & 10.565 & 367.964 \\ 
        \textbf{$^3D_3$} & 0.382 & 1.229 & 8.764 & 0.950 & 8.476 & 19.417 & 2.504 & 3.726 & 254.856 \\ \hline
    \end{tabular}}
    \label{sc}
\end{table}

\begin{table}[!ht]
    \centering
    \caption{ Optimized model parameters for channels exhibiting multi-channel scattering}
    \resizebox{\textwidth}{!}{
    \begin{tabular}{cccccccccc}
    \hline
        \textbf{States/Parameters} & \textbf{$\alpha_0$} & \textbf{$\alpha_1$} & \textbf{$\alpha_2$} & \textbf{$r_0$} & \textbf{$r_1$} & \textbf{$r_2$} & \textbf{$X_1$} & \textbf{$X_2$} & \textbf{$D_0$} \\ \hline
        \textbf{$^3S_1$} & 0.217 & 0.011 & 0.140 & 5.474 & 3.920 & 22.144 & 2.963 & 4.842 & 2.696 \\ 
        \textbf{$\epsilon$} & 0.057 & 0.411 & 0.327 & 0.142 & 0.075 & 3.902 & 0.228 & 1.227 & 42.283 \\ 
        \textbf{$^3D_1$} & 0.419 & 0.737 & 0.149 & 1.152 & 6.330 & 0.680 & 2.979 & 8.100 & 92.727 \\ \hline
    \end{tabular}}
    \label{mc}
\end{table}

\begin{table}[!ht]
    \centering
    \caption{Obtained Resonance Energy and Full Width Half Maximum for the $\alpha-d$ along with their experimental values. }
    \begin{tabular}{lllll}
    \hline
        Sates & $E_r$(Our Work) & $E_r$(Exp) & $\Gamma_{CoM}$(Our Work) & $\Gamma_{CoM}$(Exp) \\ \hline
        $^3D_1$ & 6.64 & 5.7 & 2.18 & 1.9 \\ 
        $^3D_2$ & 4.66 & 4.7 & 1.39 & 1.32 \\ 
        $^3D_3$ & 2.08 & 2.19 & 0.15 & - \\ \hline
    \end{tabular}
    \label{csp}
\end{table}

\end{document}